\documentclass{article}

\usepackage{authblk}
\usepackage{amsmath,amssymb}
\DeclareMathOperator*{\argmax}{arg\,max}

\usepackage{booktabs} % For pretty tables
\usepackage{caption} % For caption spacing
\usepackage{subcaption} % For sub-figures
\usepackage{graphicx}
\usepackage{pgfplots}
\usepackage[all]{nowidow}
\usepackage[utf8]{inputenc}
\usepackage{tikz}
\usetikzlibrary{er,positioning,bayesnet}
\usepackage{multicol}
\usepackage{algpseudocode,algorithm,algorithmicx}

\pgfplotsset{compat=1.14}

\begin{document}
\title{Traffic Signal Control and Speed Offset Coordination Using Q-Learning for Arterial Road Networks}
%
%\titlerunning{Abbreviated paper title}
% If the paper title is too long for the running head, you can set
% an abbreviated paper title here
%

\author[1]{Tianchen Yuan}
\author[2]{Petros A. Ioannou}
\affil[1,2]{CATT Laboratory, University of Southern California}
\date{}
\setcounter{Maxaffil}{0}
\renewcommand\Affilfont{\itshape\small}
%
%\authorrunning{F. Author et al.}
% First names are abbreviated in the running head.
% If there are more than two authors, 'et al.' is used.
%
%
\maketitle              % typeset the header of the contribution
\begin{abstract}

Arterial traffic interacts with freeway traffic, yet the two are controlled independently. Arterial traffic signals do not take into account freeway traffic and how ramps control ingress traffic and have no control over egress traffic from the freeway. This often results in long queues in either direction that block ramps and spill over to arterial streets or freeway lanes. In this paper, we propose an adaptive arterial traffic control strategy that combines traffic signal control (TSC) and dynamic speed offset (DSO) coordination using a Q-learning algorithm for a traffic network that involves a freeway segment and adjacent arterial streets. The TSC agent computes the signal cycle length and split based on observed intersection demands and adjacent freeway off-ramp queues. The DSO agent computes the relative offset and the recommended speeds of both ways between consecutive intersections based on their physical distance, intersection queues, and signal cycles. We evaluate the performance of the proposed arterial traffic control strategy using microscopic traffic simulations of an arterial corridor with seven intersections near the I-710 freeway. The proposed QL-based control significantly outperforms a fixed-time control and MAXBAND in terms of the travel time and the number of stops under low or moderate demands. In high-demand scenarios, the travel-time benefit provided by the QL-based control is reduced as it mitigates off-ramp and intersection queues, which is a necessary trade-off in our perspective. In addition, mutual benefit is obtained by implementing freeway and arterial traffic control simultaneously.

\end{abstract}
\section{Introduction}
The effective management of arterial traffic has become a pivotal aspect in addressing the challenges posed by growing populations and increasing vehicular densities. Traffic signal control (TSC) is the only available control option for most arterial roads nowadays. The synchronization of traffic signals facilitates the creation of green waves, allowing for more seamless progression of vehicles and reducing travel times on arterial roads. The conventional fixed-time TSC systems are proved to be insufficient in coping with the unpredictable nature of modern urban traffic \cite{wang2018review}. To address the issue, many adaptive TSC algorithms have been proposed and thoroughly studied by the research community \cite{feng2015real,jin2017group,zhang2022distributed}. The adaptive TSC system typically uses sensors and real-time data to adjust signal timings and dynamically respond to the change of traffic states. However, the benefit obtained by the adaptive signal control can be limited when the arterial roads are highly saturated due to high demands or unexpected incidents \cite{tian2008interactive,chandan2022real}. 

Although being less practiced in the real world, other traffic regulation techniques such as variable speed limit (VSL) control may also improve the arterial traffic mobility when coordinated with the traffic signal control \cite{al2014application,de2016speed}. The core idea is to control the speed of vehicles to allow as many vehicles to cross the signalized intersection during the green phase as possible, and thus, maximize the progression bandwidth. The above concept has been further extended to the study of eco-driving where the most energy-saving trajectory is determined for connected vehicles (CVs) based on the information of signal phasing and timing \cite{de2016eco,hao2018eco,yang2020eco}. Most of these studies focus on a microscopic/vehicular level and assume the signal timing is fixed.

In this paper, we aim to develop an arterial traffic control strategy that coordinates a series of traffic signals with variable speed control to improve the progression bandwidth and reduce the arterial travel time. To facilitate the coordination between different control components, we adopt the reinforcement learning (RL) framework, which has been applied in many adaptive signal control studies due to its ability to share the control policy and the perception of traffic conditions between individual agents \cite{medina2010arterial}. Besides, the RL framework can be modified to include both the signal control and the speed control as two separate agents, but maintain collaborative interactions among the agents to enhance the overall system performance.

We have designed a RL-based freeway traffic control strategy and verified its effectiveness in a mixed freeway and arterial road network in the previous study \cite{yuan2023integrated}. In this paper, we focus on the performance of the arterial part and implement a more sophisticated arterial control strategy using a similar RL framework in the same road network. In addition, we are interested in the interactions between the freeway control and the arterial control.

The contributions of this paper are summarized as follows:
\begin{enumerate}
    \item We develop an adaptive arterial traffic control strategy that coordinates traffic signal timing, offset and vehicle speed using a Q-learning (QL) algorithm. The traffic signal control (TSC) agent determines the signal cycle length and the split based on observed intersection demands and freeway off-ramp queues. The dynamic speed offset (DSO) agent determines the relative offset and the recommended vehicle speeds between two adjacent arterial signals based on the physical distance, the intersection queues and the signal cycles from the TSC agent. Both agents' reward functions aim to minimize the travel time and the queue length of freeway ramps and intersection areas.
    \item The proposed arterial traffic control improves the arterial travel time by about 15\% under low or moderate demands compared with fixed-time control, which is significantly more than the classic MAXBAND. In high-demand scenarios, the proposed control encourages queue dissipation at off-ramps and intersections so that the travel-time benefit decreases to around 10\% as an acceptable trade-off. 
    \item The proposed arterial traffic control improves the freeway travel time by around 5\% under high demands due to its ability to alleviate off-ramp queues. The freeway traffic control also benefits the proposed arterial control by producing balanced off-ramp flows and intersection demands which fit the unification of signal cycles and splits.
\end{enumerate}

The rest of the paper is organized as follows: section \ref{section:litrev} reviews relevant literature. Section \ref{section:probstate} states the problem that we aim to solve. Section \ref{section:method} presents the QL-based arterial traffic control strategy. Section \ref{section:Simu} verifies the effectiveness of the proposed approach via microscopic simulations. Section \ref{section:Conclu} concludes the paper.

\section{Related Work} \label{section:litrev}
Traffic signal control (TSC) is the most crucial and effective arterial-traffic-control approach and has been frequently studied by the transportation community for over half a century. There are two main categories of existing TSC strategies: fixed-time TSC and adaptive/traffic-responsive TSC. Fixed-time TSC operates by switching between predetermined signal programs based on the time of day, making it well-suited for stable and unsaturated traffic conditions. In 1960s, Webster and Miller laid the groundwork for modern fixed-time TSC by developing a traffic signal timing model and calculation method to minimize the average vehicle delay \cite{webster1958traffic,miller1963settings,webster1966traffic}. In \cite{calle2019computing} and \cite{yuan2023integrated}, the original Webster model has been modified to consider vehicle delays, fuel consumption and emission rates. Although the Webster model is able to optimize the performance of traffic flows for an isolated intersection, repeatedly applying the model for closely distanced signals on an arterial road may not yield the best performance. Instead it is recommended to synchronize the traffic signal timing with proper offsets to create a green wave and reduce the overall vehicle stops for red lights \cite{orcutt1993traffic}. The exact idea has been incorporated into the MAXBAND model \cite{little1981maxband}. The classic MAXBAND model lacks robustness with respect to unpredictable arterial traffic characteristics, and thus, has been modified in a number of later studies. In \cite{gartner1991multi}, Gartner et al. divided the road into multiple segments and apply MAXBAND for each segment separately. In \cite{arsava2016arterial}, Arsava et al. incorporated origin-destination (OD) information and route guidance with MAXBAND to guarantee the bandwidth allocation for major origin-to-destination flows. In \cite{de2016speed}, De Nunzio et al. combined speed advisory with MAXBAND to reduce the travel time and the energy consumption. Another widely adopted and extended fixed-time TSC scheme is TRANSYT \cite{robertson1969transyt}, which utilizes historical traffic data from the road network as input and computes the optimal signal timing via a heuristic "hill climbing" algorithm. 

The primary drawback of fixed-time TSC is the inability to manage highly saturated traffic conditions or abnormal demands due to events or incidents. To address the issue, some adaptive TSC methodologies have been investigated. In \cite{hunt1982scoot}, Hunt et al. proposed a traffic-responsive variation of TRANSYT called SCOOT, which modifies signal plans online based on observed traffic flow rates and occupancy. In \cite{mirchandani2005rhodes}, a real-time hierarchical optimized distributed effective system (RHODES) with two primary operational stages was introduced. In the first stage, the system predicts future traffic flow rates across the road network using sensor data. In the second stage, the optimal signal timing is computed with the predicted traffic flow rates from the first stage. In \cite{feng2015real}, the authors presented an adaptive phase allocation algorithm for a single signalized intersection that optimizes the phase sequence and duration using vehicle location and speed data from connected vehicles. Even though many adaptive TSC have demonstrated satisfactory performance in various field tests, a common limitation is their dependence on precise real-time traffic data and high-speed computing resources. The complexity of the internal optimization problem of some adaptive TSC is significant, which imposes restrictions on the size of arterial networks that can be effectively managed.

Reinforcement learning (RL) has been recently adopted by many researchers as a promising alternative to deal with the scalability issue in the study of adaptive TSC \cite{yau2017survey}. In \cite{medina2010arterial}, the authors proposed a basic RL structure that assigns one agent per signal to lower delays and number of stops for a two-way arterial road segment with five intersections. The RL agents deliver a more balanced distribution of the travel delay than fixed-time control. However, each RL agent optimizes the performance of a local area and the global optimum is not guaranteed. Besides, the progression bandwidth problem is not considered in \cite{medina2010arterial}. To achieve global optimum, many research efforts focused on joint state-action modeling methods where the agent learns to choose the most rewarding joint action with joint state observations \cite{kuyer2008multiagent,van2016coordinated,tan2019cooperative}. The drawback of joint state-action methods is that the number of state-action pairs grows exponentially with the number of intersections, leading to long training time and the need for a huge amount of training data for large networks. In \cite{zhang2022distributed}, the basic RL structure was upgraded to minimize the sum of intersection queue lengths and the cumulative stop delay of the entire network using a deep neural network (DNN). The authors utilized a proximal policy optimization (PPO) algorithm to strike a balance between speed and stability of training, which solves the scalability issue with global optimization in some sense. The potential concern with PPO is the high sensitivity to the choice of hyperparameters and limited exploration ability, both of which prevent the model from achieving the true optimum. 

The recent developments in the study of eco-driving reveals the great potential of vehicle speed control on improving the arterial traffic mobility \cite{de2016eco,hao2018eco,yang2020eco}. The objective of the eco-driving agent is to find a velocity trajectory for the vehicle to travel across a series of traffic signals with least stops and energy consumption. Although eco-driving algorithms are designed for vehicular-level traffic control, the concept can be implemented in flow-level traffic control, i.e. coordinating speed limits with TSC to improve the efficiency of arterial travel. 

\section{Problem Statement} \label{section:probstate}
\begin{figure}[H]
\centering
\includegraphics[width=0.9\textwidth]{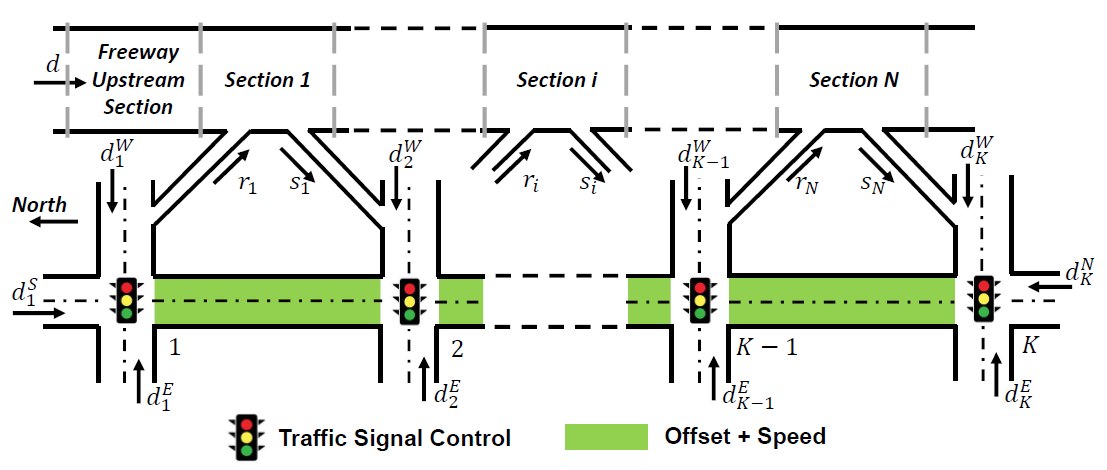}
\caption{Road network}
\label{fig:tscsa_roadnet}
\end{figure}

Consider a road network that consists of a freeway segment and the adjacent arterial streets as depicted in Fig. \ref{fig:tscsa_roadnet}. Although the freeway and arterial traffic interact frequently via ramps, they are controlled separately with very little, if any, communication. This often leads to severe traffic congestion problems such as off-ramp queue overspill where the red light queue at an arterial intersection extends all the way back to the side lane of freeway via the off-ramp. To prevent the problem, the arterial signal needs to adjust its timing to accommodate more off-ramp demands before the occurrence of overspill.

On the other hand, in spite of the evolution of traffic signal control (TSC) techniques in past decades, TSC by itself has difficulty maintaining the expected performance under high arterial traffic demands. Recent developments of eco-driving motivates us to incorporate vehicle speed control with TSC in order to increase the progression bandwidth and improve traffic mobility.

With the above considerations, we design a new arterial control strategy that coordinates traffic signal timing, offset and dynamic speed limits with the objective to reduce arterial travel time and size of queues at off-ramps and intersections. A Q-learning framework is adopted due to its ability to coordinate different control components and fast real-world implementations. The off-ramp queue is taken into account as a state variable and included in the reward function to avoid off-ramp overspill.

\section{Methodology} \label{section:method}
In this section, we provide a detailed description of the proposed Q-learning-based arterial traffic control strategy. The road network depicted in Fig. \ref{fig:tscsa_roadnet} consists of a freeway segment and an adjacent arterial corridor with multiple signalized intersections. The freeway traffic flow is characterized by a modified cell transmission model (CTM) that considers the capacity drop phenomenon and bounded accelerations \cite{yuan2023integrated}. Each CTM section contains at least one ramp and each ramp connects to the East leg of an arterial intersection. Note that some ramp connections are omitted in Fig. \ref{fig:tscsa_roadnet} due to limited drawing space. The proposed arterial traffic control is operated by two agents: a traffic signal control (TSC) agent that computes the signal plan (with zero offset) for each intersection and a dynamic speed offset (DSO) agent that determines the proper signal offset and vehicle speed between two intersections.

\subsection{TSC Agent}
\begin{figure}[H]
\centering
\includegraphics[width=0.6\textwidth]{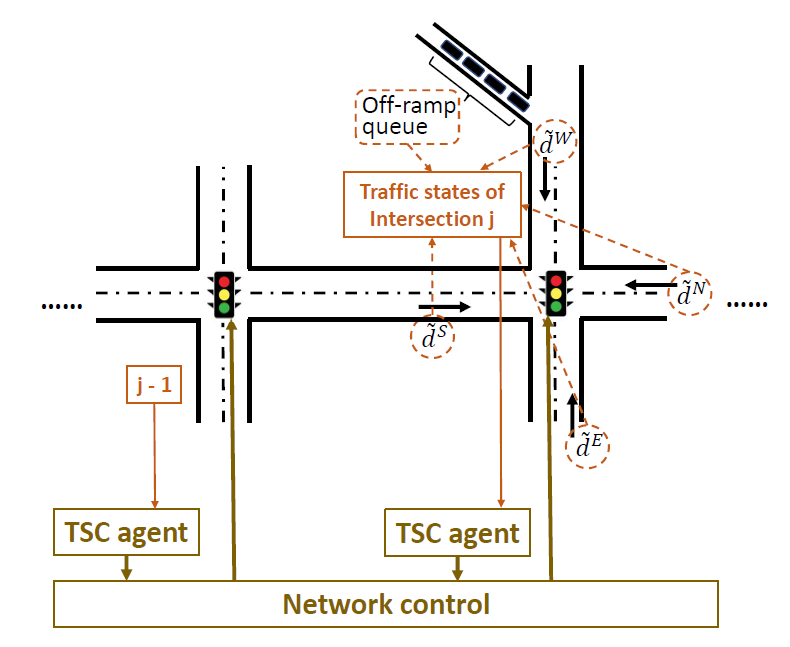}
\caption{Traffic signal control (TSC) agent}
\label{fig:qltsc}
\end{figure}

The TSC agent computes the signal plan which includes the cycle length and phase split for each arterial signal based on the observations of traffic states as shown in Fig. \ref{fig:qltsc}. We introduce a network-level controller to receive the outputs of all the TSC agents and generate final signal plans. The network control unifies the signal plan for all intersections so that the progression bandwidth on the horizontal arterial in Fig. \ref{fig:tscsa_roadnet} can be maximized with the assistance of the DSO agent. This unification mechanism can be either activated or deactivated. The control cycle is set to 5 min, which means the signal plan is updated every 5 min during the simulation. To apply QL algorithms, we first introduce the definition of states, actions and reward function for the TSC agent.

\subsubsection{States:}
The states of the TSC agent are defined as
\begin{equation} \label{eq:TSCstate}
    X_{ts} = [w_s,\Tilde{d}^S,\Tilde{d}^E,\Tilde{d}^N,\Tilde{d}^W]
\end{equation}
where $w_s$ is the queue length of the off-ramp that is connected to the freeway side of the corresponding intersection, and $\Tilde{d}^S,\Tilde{d}^E,\Tilde{d}^N,\Tilde{d}^W$ are measured incoming demands of the intersection from four directions in vehicles per hour (veh/h), with the superscript denoting the direction (e.g. $S$ for Southbound). These traffic states are measured every 30 seconds. The demand is estimated as the average measured flow rate of the upstream link in the past control cycle. $w_s$ is set to 0 at the beginning of each simulation, and then calculated as the average observed off-ramp queue length in the past control cycle.

We use $S_w$ to denote the state space of $w_s$ and $S_d$ to denote the state space of $\Tilde{d}^*$ ($*$ stands for the direction). To reduce the problem dimension and improve the training efficiency, we discretize the continuous state space so that $S_w = \{0,50,100,...,500\}$ and $S_d = \{0,100,200,...,4000\}$. Note that $500$ m is slightly larger than the length of the longest off-ramp and $4000$ veh/h is larger than the maximal historical link flow. Therefore we choose these two values as the upper limit of the discrete state spaces.  

\subsubsection{Actions:}
The actions executed by the TSC agent are expressed as
\begin{equation} \label{eq:TSCaction}
    A_{ts} = [T_c,g_1,g_2]
\end{equation}
where $T_c$ is the cycle length, which is the sum of green time of all the phases plus the lost time during phase transitions. $g_1,g_2$ are phase-split ratios to be used to compute the green time of each phase, and below we will present the exact expressions used to calculate them. 

\begin{figure}[H]
\centering
\includegraphics[width=0.9\textwidth]{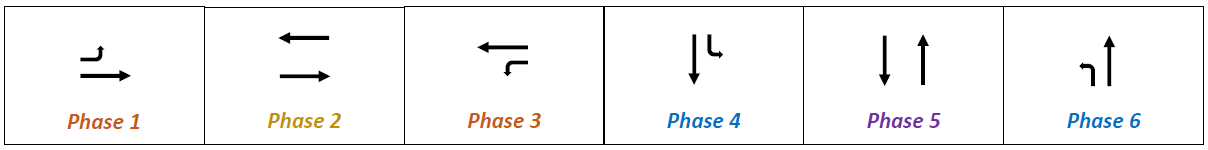}
\caption{Traffic signal phases}
\label{fig:6phase}
\end{figure}

The default phase scheme involves 6 phases as depicted in Fig. \ref{fig:6phase}. 
The green time of each phase is denoted as $T_{g,j}$ where $j=1,2,...,6$ is the phase index. To simplify the phase split computation, we assume the phase with the same color in Fig. \ref{fig:6phase} has the same green time, i.e. $T_{g,1} = T_{g,3}$, $T_{g,4} = T_{g,6}$. In addition, $T_{g,1}/T_{g,2}=T_{g,4}/T_{g,5}$.

With the above assumptions, we only need two ratios to compute the green time of each phase - $g_1$
is the green time of first 3 phases divided by the green time of all 6 phases, $g_2$ is the green time of the first phase divided by the green time of first 3 phases, i.e.
\begin{equation} \label{eq:phasesplit_ratios}
    \begin{aligned}
    g_1 &= \frac{T_{g,1}+T_{g,2}+T_{g,3}}{T_{g,1}+T_{g,2}+T_{g,3}+T_{g,4}+T_{g,5}+T_{g,6}} \\
    g_2 &= \frac{T_{g,1}}{T_{g,1}+T_{g,2}+T_{g,3}} 
    \end{aligned}
\end{equation}
Note that the green time of all 6 phases is not equal to the cycle length $T_c$, instead we have 
\begin{equation}
    T_{g,1}+T_{g,2}+T_{g,3}+T_{g,4}+T_{g,5}+T_{g,6} = T_c - T_l
\end{equation}
where $T_l$ is the loss time, defined as the time period during which no vehicles pass through the intersection due to phase transitions.

The action space of the cycle length $A_c$ is set to $\{40,50,60,...,180\}$ in seconds according to \cite{yuan2023integrated}. The action space of $g_1$ is $A_{g_1}=\{0.2,0.3,0.4,...,0.8\}$. The action space of $g_2$ is $A_{g_2}=\{0.1,0.2,0.3,0.4\}$. After choosing $g_1$ and $g_2$, the phase split can be computed as follows:
\begin{equation} \label{eq:phasesplit_comp}
    \begin{aligned}
    T_{g,1} &= T_{g,3} = (T_c - T_l)g_1g_2\\
    T_{g,2} &= (T_c - T_l)g_1(1-2g_2) \\
    T_{g,4} &= T_{g,6} = (T_c - T_l)(1-g_1)g_2\\
    T_{g,5} &= (T_c - T_l)(1-g_1)(1-2g_2)
    \end{aligned}
\end{equation}

\subsubsection{Reward:}
\begin{figure}[H]
\centering
\includegraphics[width=0.8\textwidth]{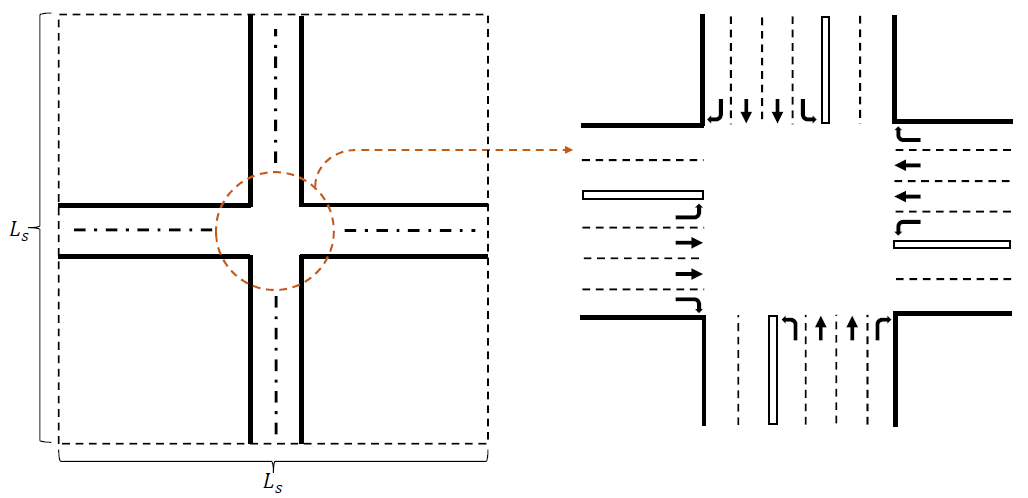}
\caption{The square area for TSC to compute the average travel time}
\label{fig:inter_config}
\end{figure}

The first objective of the traffic signal control is to reduce the average travel time for a square area centered at the intersection as depicted in Fig. \ref{fig:inter_config}.
The side length of the square area $L_s$ is chosen to be long enough to measure and monitor the queues at the intersection. In this paper, we set $L_s=400$ m.

The average travel time $T_t$ is computed as
\begin{equation} \label{eq:att}
    T_t = \frac{1}{N_v}\sum_{i=1}^{N_v}(t_{i,out}-t_{i,in})
\end{equation}
where $N_v$ is the number of vehicles traveling through the square area during the last control cycle, $t_{i,in}$ and $t_{i,out}$ is the time vehicle $i$ enters and exits the square area. 

The TSC agent also interacts with the off-ramp queue. An effective reward function should exhibit a negative correlation with the average travel time $T_t$ and the average off-ramp queue length $w_s$. Considering the above requirements, the reward function is defined as
\begin{equation} \label{eq:TSCreward}
    R_{ts} = \max\{0,1-\frac{w_s}{w_s^r}\} \cdot \frac{L_s}{T_t v_a}
\end{equation}
where $w_s^r$ is the reference off-ramp queue length that we do not want $w_s$ to reach. $v_a$ is the default arterial speed limit. $w_s^r$ should be slightly less than the off-ramp queue capacity to prevent the overspill. In the simulations, we set $w_s^r=400$ m and $v_a=60$ km/h.

$R_{ts}$ has a maximum value of 1, which is obtained when there is no queue at the off-ramp and the average travel time $T_t=L_s/v_a$. This is a rare case which occurs when there is no travel delay at the intersection. $R_{ts}$ reaches the minimum value of 0 when $w_s$ exceeds the reference value $w_s^r$, which effectively prevents off-ramp queue overspill. In general, a higher $R_{ts}$ reflects less off-ramp queue length and average travel time. Note that the objective of Q-learning is not only to maximize the immediate reward defined in (\ref{eq:TSCreward}), but to maximize a cumulative long-term reward where the reward of each time step is computed by (\ref{eq:TSCreward}).

\subsubsection{Network Control:}
The network controller receives the intended actions from all TSC agents within the arterial network and selects a unified action for each action variable of all signals based on a majority rule as follows: 
\begin{itemize}
    \item If the most common action is the intended action of more than half TSC agents, it selects it as the unified action for all signals.
    \item If the most common action is supported by half TSC agents or less, it computes the average of all intended actions and selects the common action to be the one that is closest to the average value. If the average value lies in the middle of two actions exactly, it selects the larger one.
\end{itemize}

To demonstrate the majority rule proposed above, we provide an example of an arterial corridor with four signalized intersections and discuss three cases as follows:
\begin{itemize}
    \item Case 1: the intended signal cycles are $\{60, 60, 60, 70\}$s. Since more than half agents favors 60s, the network controller selects 60s as the common cycle.
    \item Case 2: the intended signal cycles are $\{60, 60, 70, 80\}$s. No action is favored by more than half agents. The average cycle length is 67.5s. The network controller selects the cycle that is closest to 67.5s, i.e. 70s.
    \item Case 3: the intended signal cycles are $\{60, 60, 70, 70\}$s. No action is favored by more than half agents. The average cycle length is 65s, which lies in the middle of two available options - 60s and 70s. The network controller selects the larger one, i.e. 70s.
\end{itemize}

Although the unification may compromise the travel time of some particular intersection areas, it benefits the overall arterial travel time and traffic mobility in most circumstances with the collaboration of the DSO agent. The unification mechanism is deactivated during the training process so that more state-action pair can be visited.

\subsection{DSO Agent}
The DSO agent searches for the optimal offset and speed recommendations between two adjacent arterial signals based on the observations of signal plans, intersection queue length and link distance as shown in Fig. \ref{fig:qlsa}. The control cycle is set to 5 min to be consistent with the TSC agent. The states, actions and reward function for the DSO agent are described as follows:

\begin{figure}[H]
\centering
\includegraphics[width=0.6\textwidth]{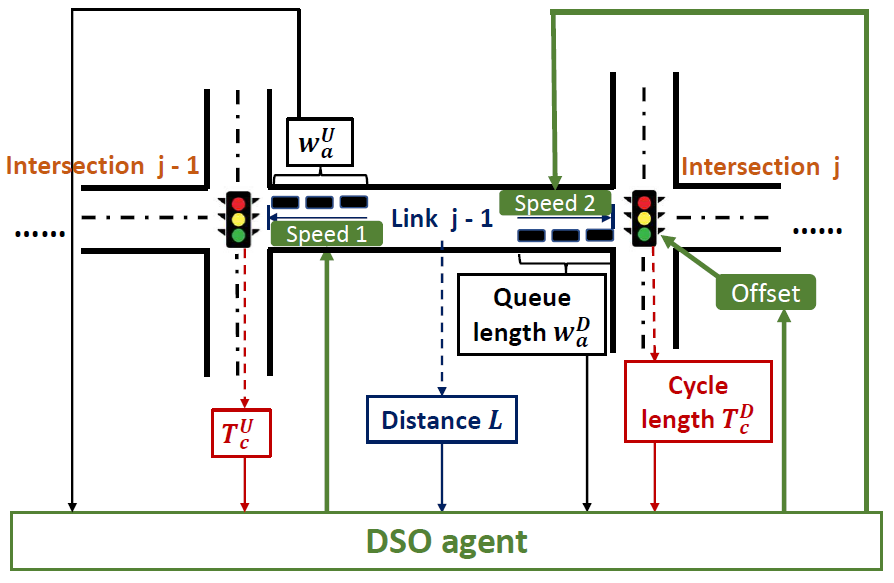}
\caption{Dynamic speed offset agent}
\label{fig:qlsa}
\end{figure}

\subsubsection{States:}
The states of the DSO agent are defined as
\begin{equation} \label{eq:SAstate}
    X_{so} = [T_c^U,T_c^D,w_a^U,w_a^D,L]
\end{equation}
where $T_c^U$ is the cycle length of the upstream signal, $T_c^D$ is the cycle length of the downstream signal, $w_a^U$ is the queue length at the Northbound approach of the upstream intersection, $w_a^D$ is the queue length at the Southbound approach of the downstream intersection, and $L$ is the link distance.

$T_c^U$ and $T_c^D$ are determined by the TSC agent after each control cycle. $w_a^U$ is observed every 30 seconds and we take the average of all observed values of $w_a^U$ during each control cycle as its state value. The same applies to $w_a^D$. The link distance $L$ is assumed known for each pair of adjacent intersections within the network. 

The state space of each state variable is also discretized to accelerate the training process. The state space of $T_c^U$ and $T_c^D$, denoted as $S_c$, is the same as the action space of the cycle length $A_c$, i.e. $S_c=\{40,50,60,...,180\}$ in seconds. The state space of $w_a^U$ and $w_a^D$, denoted as $S_{w_a}$, is set to $\{0,50,100,...,250\}$ in meters. The state space of $L$, denoted as $S_L$, is defined as $\{1000,1100,1200,...,2500\}$ in meters. 

\subsubsection{Actions:}
The actions executed by the DSO agent are expressed as
\begin{equation} \label{eq:SAaction}
    A_{so} = [T_o,v_r^D,v_r^U]
\end{equation}
where $T_o$ is the offset of signal $j$ with respect to signal $j-1$ expressed in seconds, which means that the cycle initiation of signal $j$ is later than signal $j-1$ by $T_o$ seconds. $v_r^D$ is the recommended speed for downstream traffic in link $j-1$, and $v_r^U$ is the recommended speed for upstream traffic in link $j-1$.

It is unnecessary for the offset $T_o$ to be larger than the cycle length of signal $j-1$ \cite{eom2020traffic}, and thus, the action space of $T_o$, denoted as $A_o$, is defined as $\{0,5,10,...,T_c'-5\}$ in seconds, where $T_c'$ is the cycle length of the upstream signal. The action space of $v_r^D$ and $v_r^U$, denoted as $A_{v_r}$, is set to $\{40,45,50,...,80\}$ in km/h. 

\subsubsection{Reward:}
The task for the DSO agent is to reduce the average travel time on link $j-1$ and the queue lengths at both intersections by setting a proper offset and speed recommendations. Note that higher speed does not necessarily reduce the travel time as the vehicle may run into the red phase at the signal. In that case, we prefer to let the vehicle approach the signal with lower speed so that it passes directly under the green light and does not join the red light queue.

Guided by the above ideas, we define the reward function for the DSO agent as follows:
\begin{equation} \label{eq:SAreward}
    R_{so} = \max\{0,1-\frac{w_a^U}{w_a^r}\} \cdot \max\{0,1-\frac{w_a^D}{w_a^r}\} \cdot \frac{3L}{4T_t' v_a}
\end{equation}
where $w_a^r$ is the reference queue length of an intersection approach and $T_t'$ is the average travel time for both directions on link $j-1$. $T_t'$ is calculated using (\ref{eq:att}). In the simulations, we set $w_a^r=200$ m.

$R_{so}$ reaches the maximum value of 1 if all vehicles pass through the link at maximum possible speed (80 km/h) and the two approaches have no queue at all. The minimum value of $R_{so}$ is 0, which is obtained when either queue at the two approaches exceeds the reference value $w_a^r$. A larger value of $R_{so}$ reflects less intersection queue length and link travel time. Note that the objective of Q-learning is to maximize a cumulative long-term reward where the reward of each time step is computed by (\ref{eq:SAreward}).

\subsection{Q-Learning}
A crucial factor to enhance the effectiveness of the proposed arterial traffic control method is the coordination between the TSC agent and the DSO agent. Most rule-based and simulation-based algorithms have each control component accomplish its own objective without any coordination \cite{dion2002rule,ekeila2009development,chandan2017real}. The use of optimization framework may provide a solution that facilitates the coordination as multiple sub-controllers working toward a common objective function \cite{lin2004enhanced,feng2015real,de2016speed}. However, the optimization problem formulation for the road network under consideration is overly complex considering the network size and the mixed freeway and arterial environment, which leads to very time-consuming computations which limit its practicality. A more computationally feasible approach that also takes the coordination and optimality into account is reinforcement learning (RL) \cite{medina2010arterial,jin2017group,zhang2022distributed}. 

During the RL training process, the agent learns the best policy $\pi$ that maximizes the total expected discounted rewards $V_\pi(x)$ for each state $x$ via trial and error. Note that $V_\pi(x)$ is a long-term return and should not be confused with the immediate reward of both agents defined in (\ref{eq:TSCreward}) and (\ref{eq:SAreward}). $V_\pi(x)$ is expressed as follows
\begin{equation} \label{eq:vpi0}
    V_\pi(x) = \mathbb{E}_\pi[\sum_{k=0}^{\infty}\gamma^k R_{k}|x]
\end{equation}
where $x$ is the current state defined in (\ref{eq:TSCstate}) and (\ref{eq:SAstate}), $\pi$ is a policy that suggests which action should be taken for each possible state, $k=0$ is the current time step, $R_k$ is the reward defined in (\ref{eq:TSCreward}) and (\ref{eq:SAreward}) at time step $k$, $\gamma \in [0,1)$ is the discount factor that determines the importance of the future reward. A $\gamma$ close to 0 makes the agent short-sighted (only considering current rewards), while a $\gamma$ close to 1 makes it far-sighted (considering future rewards). The future reward gets more and more discounted as it steps away from the current time.

Without loss of generality, we assume the states of both TSC and DSO agents are memoryless, which means their future states only depend on the current states and actions. Therefore, searching the optimal policy $\pi$ can be formulated as a Markov Decision Process (MDP) problem and solved by the Bellman equation \cite{barron1989bellman,sutton2018reinforcement}. To express $V_\pi(x)$ using the Bellman equation, we first split it into the immediate reward and the discounted future rewards
\begin{equation} \label{eq:vpi1}
    V_\pi(x) = \mathbb{E}_\pi[R_0 + \gamma \sum_{k=1}^{\infty}\gamma^{k-1} R_{k}|x]
\end{equation}
The term $\sum_{k=1}^{\infty}\gamma^{k-1} R_{k}$ is the sum of discounted future rewards starting from the next time step $k=1$, which is essentially $V_\pi(x')$ where $x'$ is the state at $k=1$. Therefore, we can rewrite $V_\pi(x)$ as
\begin{equation} \label{eq:vpi2}
    V_\pi(x) = \mathbb{E}_\pi[R_0 + \gamma V_\pi(x')|x]
\end{equation}

Then we expand the expectation form by considering the policy $\pi$ and transition probability to derive the Bellman equation \cite{barron1989bellman} for $V_\pi(x)$ as
\begin{equation} \label{eq:vpi_bellman}
    V_\pi(x) = \sum_{a}\pi(a|x)   \sum_{x'}P_a(x,x')(R_a(x,x') + \gamma V_\pi(x'))
\end{equation}
where $\pi(a|x)$ is the probability of taking action $a$ in the current state $x$ under policy $\pi$, $P_a(x,x')$ is the transition probability from $x$ to the future state $x'$ via action $a$, and $R_a(x,x')$ is the reward obtained from the above transition. $x,a,R$ are the states, actions and reward specified by (\ref{eq:TSCstate})(\ref{eq:TSCaction})(\ref{eq:TSCreward}) for the TSC agent and (\ref{eq:SAstate})(\ref{eq:SAaction})(\ref{eq:SAreward}) for the DSO agent respectively. 

$V_\pi(x)$ in (\ref{eq:vpi_bellman}) is an iterative term and its optimal value can be potentially solved using dynamic programming (DP). However, the transition probability $P_a(x,x')$ is unknown in the traffic environment, which motivates us to adopt model-free techniques such as Q-learning (QL) \cite{sutton2018reinforcement}. The Q-value $Q(x,a)$ is defined as the total expected discounted rewards after we take action $a$ in state $x$. $Q(x,a)$ directly tells us the expected return without needing a model to calculate it for each action. 

According to the definition of $Q(x,a)$ and the Markov property, the optimal $V_\pi (x)$ is derived by selecting the best action at each state
\begin{equation} \label{eq:vpi_q}
    V_\pi^*(x) = \max_a Q^*(x,a)
\end{equation}
Therefore, searching the optimal policy $\pi$ is equivalent to finding the action that maximizes the Q-value for each state
\begin{equation} \label{eq:optpolicy}
    \pi^*(x) = \argmax_a Q^*(x,a)
\end{equation}
Plugging the above optimal policy into (\ref{eq:vpi_bellman}), we have
\begin{equation}
    Q^*(x,a) = \sum_{x'}P_a(x,x')(R_a(x,x') + \gamma \max_{a'}Q^*(x',a'))
\end{equation}
Equivalently
\begin{equation} \label{eq:q_iter_TP}
    Q^*(x,a) = R_a(x,x') + \gamma \sum_{x'}P_a(x,x')\max_{a'}Q^*(x',a')
\end{equation}
where $\max_{a'}Q^*(x',a')$ represents the maximum possible future return after moving to the future state $x'$ and taking the best action $a'$. 

The Q-value iteration equation (\ref{eq:q_iter_TP}) cannot be solved since the transition probability is unknown. However, it indicates that the optimal Q-value for each state-action pair should converge to $R_a(x,x') + \gamma \max_{a'}Q^*(x',a')$, according to which the following Q-value update equation is motivated \cite{sutton2018reinforcement}.
\begin{equation} \label{eq:q_update}
    Q(x,a) \leftarrow Q(x,a) + \eta[R(x,a) + \gamma \max_{a'} Q(x',a') - Q(x,a)]
\end{equation}
where $\eta$ is the learning rate that determines to what extent the newly acquired information will override the old information. $R(x,a) + \gamma \max_{a'} Q(x',a')$ is the newly acquired estimate of Q-value which consists of both the immediate reward and the discounted future rewards. The discount factor $\gamma$ is selected as 0.9 \cite{goodfellow2016deep}. We consider the convergence of $Q(x,a)$ is achieved when the difference between the updated $Q(x,a)$ and the previous $Q(x,a)$ is less then 0.01. 

In the case of TSC agents, the state $x$ refers to $X_{ts}$ specified by (\ref{eq:TSCstate}). The action $a$ refers to $A_{ts}$ specified by (\ref{eq:TSCaction}). The immediate reward $R(x,a)$ is computed by (\ref{eq:TSCreward}). In the case of DSO agents, the state $x$ refers to $X_{so}$ specified by (\ref{eq:SAstate}). The action $a$ refers to $A_{so}$ specified by (\ref{eq:SAaction}). The immediate reward $R(x,a)$ is computed by (\ref{eq:SAreward}).

The learning rate $\eta$ is a crucial parameter in QL algorithms. A commonly practiced training approach is to gradually reduce $\eta$ during the process so that the convergence of the Q-value is achieved with reasonable efficiency. Note that a simple time-dependent decrease of $\eta$ is not effective, as different states and actions are visited at different stages of learning. Instead, a specific $\eta$ should be assigned to each state-action pair, and its value decreases each time the pair is visited \cite{li2017reinforcement}. Therefore,
\begin{equation} \label{eq:Qlearningrate}
    \eta(x,a) = \biggr[\frac{1}{1+n(x,a)(1-\gamma)}\biggr]^{0.6}
\end{equation}
where $n(x,a)$ is the number of times the state-action pair $(x,a)$ has been visited.

To maintain the balance between exploration and exploitation in the learning process, we adopt an action selection strategy based on the Softmax function as follows \cite{zhao2011dhp}
\begin{equation} \label{eq:QlearningSoftmax}
    p(a|x) = \frac{\exp{(Q(x,a)/T)}}{\sum_{b \in A}\exp{(Q(x,b)/T)}} 
\end{equation}
where $p(a|x)$ is the probability of selecting action $a$ at state $x$, and $T$ is the temperature. $T$ is a hyperparameter that controls the level of exploration. Lower $T$ makes the policy more deterministic, favoring exploitation, while higher $T$ increases exploration by making the policy more stochastic. In our design, we want both agents to favor exploitation slightly more than exploration and thus select $T=0.5$ \cite{goodfellow2016deep}.

\subsection{Training Process}
\begin{figure}[H]
\centering
\includegraphics[width=0.8\textwidth]{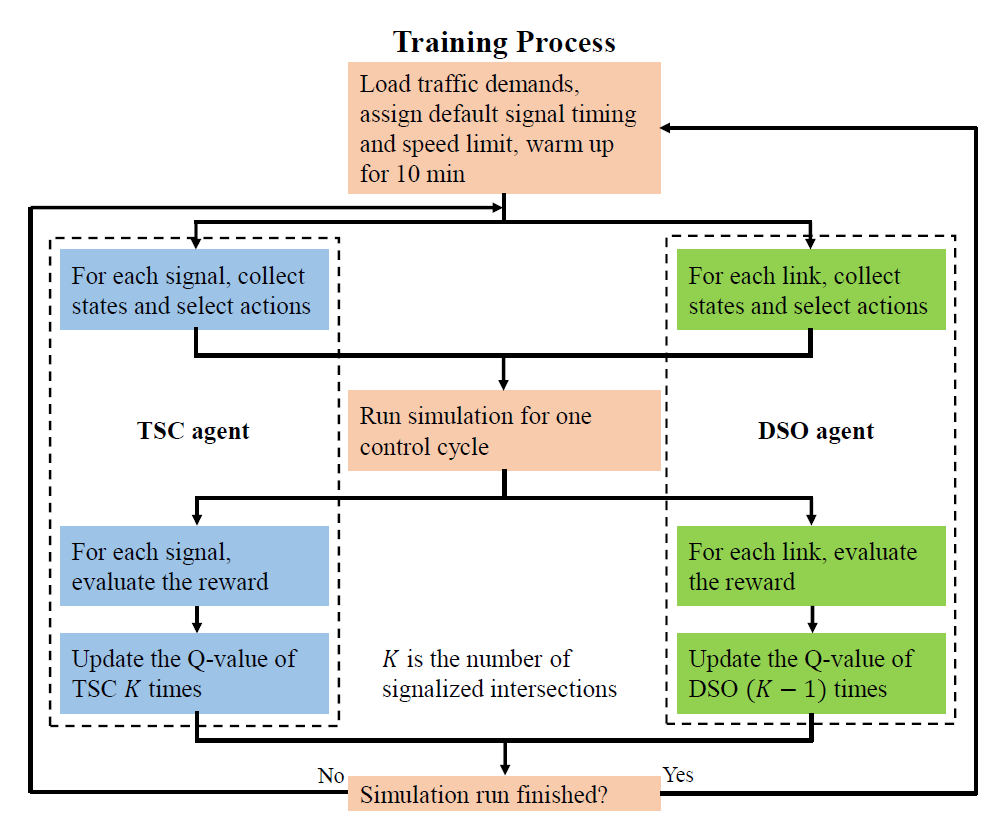}
\caption{Training process}
\label{fig:training_proc}
\end{figure}

The training process depicted in Fig. \ref{fig:training_proc} is carried out using microscopic simulation software PTV VISSIM 10 in a simulation network that will be illustrated in section \ref{section:Simu}. The freeway demands are generated using hourly traffic volume data from the Caltrans Performance Measurement System (PeMS) in April 2019. The arterial demands are generated using hourly traffic counts from LADOT Database in April 2019. The Q-value for each state-action pair is initialized randomly using a uniform distribution on $[0,1]$ to encourage exploration \cite{sutton2018reinforcement}. This range is consistent with the range of the rewards of both agents so that overshooting can be effectively avoided.

As shown in Fig. \ref{fig:training_proc}, the simulator starts by loading corresponding traffic demand at each entrance of the road network and assigning default signal plans and speed limits. The training of both TSC and DSO agents begins after a 10-min warm-up. At the beginning of each control cycle, the TSC agent first collects the state $X_{ts}$ for each signal, and selects an action $A_{ts}$ using (\ref{eq:QlearningSoftmax}) based on current Q-values. The network control is deactivated during the training process to encourage the exploration, and the TSC action of different signals may vary. At the end of the control cycle, a reward value is computed using (\ref{eq:TSCreward}) for each signal. The new state $X_{ts}'$ is collected. Then we update the Q-value for the state-action pair $(X_{ts},A_{ts})$ using (\ref{eq:q_update}). Since there are $K$ signals within the arterial network, the Q-values are updated $K$ times in total after one control cycle. The training of the DSO agent follows a similar procedure except its Q-values are updated $(K-1)$ times due to the number of links being $(K-1)$. 

The training process completes when $Q(x,a)$ converges for all state-action pairs of both agents. Considering the restrictions of VISSIM, each training simulation run lasts for 12 hours and the whole training process may take hundreds of simulation runs to complete. To fully utilize the traffic data, the 12-hour window of each simulation run shifts afterward by 1 hour from the previous run. For example, if the current run uses the traffic data from 6 a.m. to 6 p.m. of a specific day, the next run will use the traffic data from 7 a.m. to 7 p.m. of the same day.

The arterial incident is not considered in this paper. However, a freeway incident is introduced with two purpose: the first one is to bring new traffic conditions into the training to encourage the exploration of more state-action pairs by both agents; the second purpose is to activate the full version of freeway traffic control which consists of variable speed limit (VSL), lane change (LC) and ramp metering (RM), so that the agents in the arterial network can be trained to work with the freeway control. During the training, the incident on freeway has a probability of 0.25 to occur at the beginning of each hour and is cleared in 20 minutes.

\section{Experimental Study} \label{section:Simu}
\subsection{Simulation Network and Parameters} \label{subsec:simnetandpara}
\begin{figure}[H]
\centering
\includegraphics[width=0.9\textwidth]{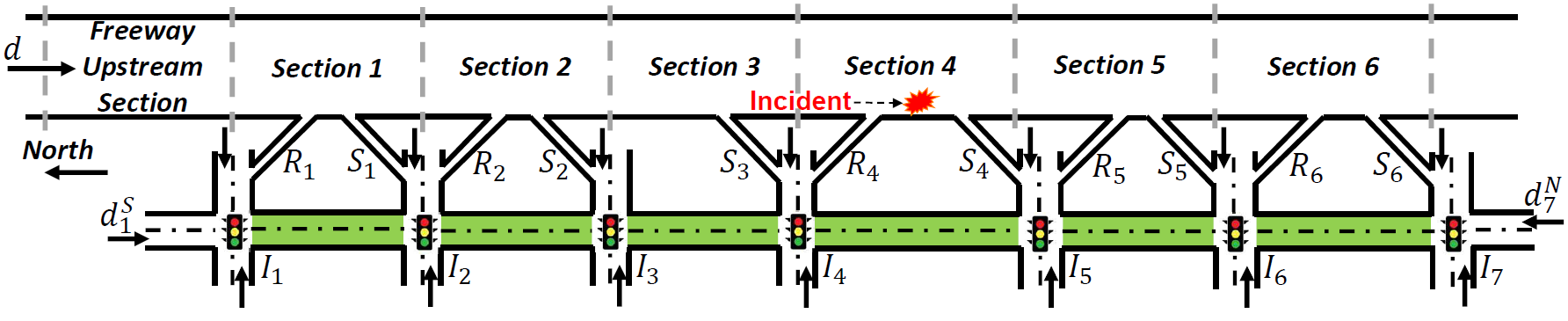}
\caption{Simulation road network}
\label{fig:sim_roadnet}
\end{figure}

\begin{figure}[H]
\centering
\includegraphics[width=0.9\textwidth]{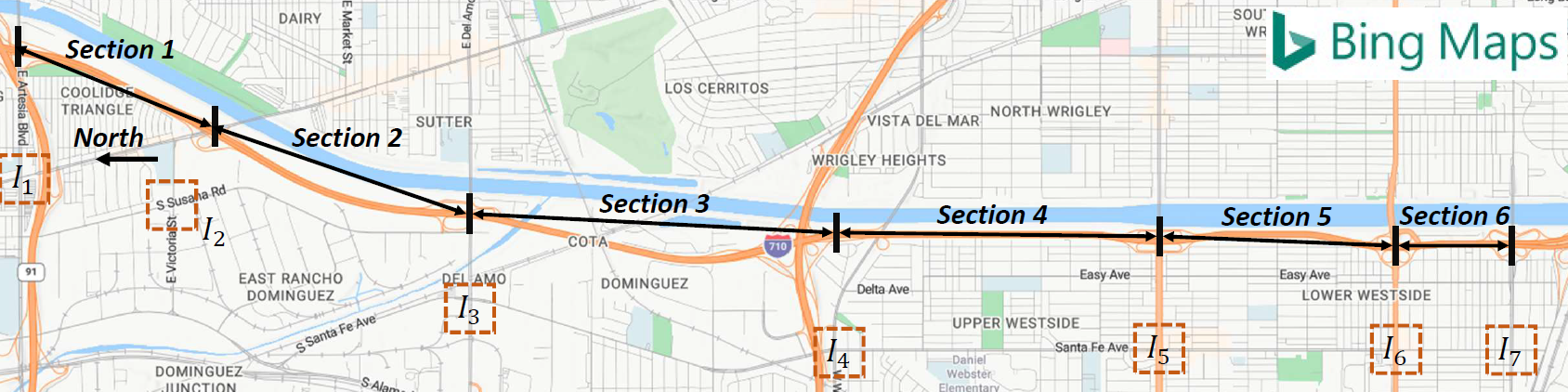}
\caption{Simulation road network on Bing Map}
\label{fig:sim_bingmap}
\end{figure}

We implement the proposed arterial traffic control strategies with the freeway traffic control strategies designed in \cite{yuan2023integrated} simultaneously and perform microscopic simulations using the commercial software PTV Vissim 10 over a mixed freeway and arterial road network which is depicted in Fig. \ref{fig:sim_roadnet}. The network consists of a 16-km segment of the I-710 freeway and an adjacent arterial corridor with 7 signalized intersections in Los Angeles, California, United States. The freeway segment is divided into 6 CTM sections and one upstream section, where the freeway controller deploys the first VSL sign dynamically \cite{yuan2022selection}. There are 5 on-ramps and 6 off-ramps that connect the freeway segment with the arterial region. The real-world locations of 7 arterial intersections are marked by orange squares in Fig. \ref{fig:sim_bingmap}. The selected intersections are major ones whose traffic states are strongly correlated with the traffic states of neighboring freeway sections. To simplify the simulation network, some minor arterial streets and vertical freeways are ignored, and the selected intersections are directly linked.

Traffic demands are produced at one freeway entrance and 16 arterial entrances as implied by arrows in Fig. \ref{fig:sim_roadnet}. The freeway demands are generated using hourly traffic volume data from the Caltrans Performance Measurement System (PeMS) in April 2019. The arterial demands are generated using hourly traffic counts from LADOT Database in April 2019. The turning ratio at an intersection approach or a ramp is determined by the ratio of the historical average traffic counts of each direction based on the LADOT data. The incident indicated in Fig. \ref{fig:sim_roadnet} triggers a side-lane-closure on freeway and potentially impacts both freeway and arterial traffic. To evaluate the trained agents, three levels of traffic demands are considered: a low-demand level based on 1-3 a.m. weekday traffic, a moderate-demand level based on 12-2 p.m. weekday traffic, and a high-demand level based on 5-7 p.m. weekday traffic. Each evaluation simulation run lasts for 40 min. In the presence of an incident, it occurs after a 10-minute warm-up and is cleared at 30 minutes.

With three demand levels and the incident occurrence option, we have a total of six evaluation scenarios. The interested control strategies to be evaluated under these scenarios are listed as follows:
\begin{itemize}
    \item No freeway control (NFC): the available freeway control components (variable speed limit, lane change, ramp metering) are all inactive.
    \item Integrated freeway control (IFC): a freeway control strategy that coordinates all the control components, proposed in \cite{yuan2023integrated}.
    \item Fixed-time arterial traffic signal control (FAC): each signal has a fixed cycle of 120 s, a fixed split and zero offset. The arterial speed limit is fixed to 60 km/h.
    \item MAXBAND: a classic arterial signal control algorithm that maximizes the progression bandwidth \cite{little1981maxband}. 
    \item QL-based arterial traffic control without unification (QAC): the proposed arterial traffic control strategy without unifying the cycle and split.
    \item QL-based arterial traffic control with unification (QACU): the proposed arterial traffic control strategy with active unification logic.
\end{itemize}

\subsection{Evaluation Criteria}
We evaluate the performance of a few combinations of the above control strategies for each interested scenario. Note that the warm-up period is excluded in any type of evaluation. The evaluation criteria are listed as follows \cite{zhang2016combined,yuan2023integrated}:
\begin{itemize} 
    \item Freeway average travel time ($T_t^f$): the average time spent for each vehicle to travel through the freeway segment. The computation of $T_t^f$ follows (\ref{eq:att}). Vehicles that enter or exit from ramps are excluded. 
    \item Arterial average travel time ($T_t^a$): the average time spent for each vehicle to travel through the entire arterial region. The computation of $T_t^a$ follows (\ref{eq:att}). Only vehicles that have travelled from intersection 1 ($I_1$) to intersection 7 ($I_7$) through the arterial road are counted. 
    \item Arterial average number of stops ($\bar{s}$): the average number of stops performed by each vehicle when traveling through the entire arterial region.
    \begin{equation}
        \bar{s} = \frac{1}{N_v}\sum_{i=1}^{N_v} s_i
    \end{equation}
    where $s_i$ is the number of stops performed by vehicle $i$. Only vehicles that have travelled from $I_1$ to $I_7$ through the arterial road are counted.
    \item Arterial average emission rates of CO$_2$ ($E$): calculated using the MOVES model proposed by the Environment Protection Agency \cite{epa2010motor}. 
    \begin{equation}
        E = \sum_{i=1}^{N_v} E_i / \sum_{i=1}^{N_v} l_i
    \end{equation}
    where $E_i$ is the emission produced by vehicle $i$ and $l_i$ is the travelled distance of vehicle $i$. Only vehicles that have travelled from $I_1$ to $I_7$ through the arterial road are counted.
    \item Average off-ramp queue length ($\bar{w}_s$):
    \begin{equation}
        \bar{w}_s = \sum_{i=1}^{N} \bar{w}_{s,i} / N_s
    \end{equation}
    where $N$ is the number of freeway sections, $\bar{w}_{s,i}$ is the average queue length of off-ramp $i$ during the simulation, $N_s$ is the number of off-ramps.
    \item Average queue length of arterial intersections ($\bar{w}_a$):
    \begin{equation}
        \bar{w}_a = \sum_{k=1}^{K} (\bar{w}^N_k + \bar{w}^S_k + \bar{w}^E_k + \bar{w}^W_k) / 4K
    \end{equation}
    where $K$ is the number of arterial intersections, $\bar{w}^N_k$ is the average queue length of the Northbound approach of intersection $k$ during the simulation.
\end{itemize}

\subsection{Evaluation Results}
We present the evaluation results of six interested scenarios in six tables respectively. Each value in these tables is the average of ten random simulation runs. There are two freeway control strategies and four arterial control strategies to be evaluated as illustrated in section \ref{subsec:simnetandpara}. However, some options may not be necessary for some specific scenarios. For instance, the integrated freeway control (IFC) is not needed when there is no incident on freeway since the freeway demand is always within its normal capacity. Thus, we only focus on representative combinations of freeway and arterial control strategies for the sake of simplicity.

Table \ref{tb:eval_lowD} presents the evaluation results of a low-demand scenario without the freeway incident. In table \ref{tb:eval_lowD}, the freeway travel time $T_t^f$ is not affected by the variation of arterial control since the freeway traffic always follow the free-flow speed without any restriction. The fixed-time policy (FAC) serves as the reference method and delivers the worst performance under all arterial performance measurements. The percentages in brackets quantify the performance improvement by implementing the corresponding control scheme versus FAC. The QL-based control (QAC and QACU) performs better than the MAXBAND, especially in terms of the travel time and the number of stops. The unification logic also slightly improves the overall performance as we compare the results of QAC with QACU.

\begin{table}[h]
\begin{center}
\setlength{\tabcolsep}{6pt} % Default value: 6pt
\caption{Low-Demand without Incident} \label{tb:eval_lowD}
\begin{tabular}{|c|c|c|c|c|}
\hline
Freeway control & \multicolumn{4}{|c|}{NFC}    \\     \hline
Arterial control & FAC & MAXBAND & QAC & QACU \\    \hline
$T_t^f$(s) & 634 & 634 & 633 & 633              \\
$T_t^a$(s) & 968 & 853(12\%) & 816(16\%) & 800(17\%)    \\
$\bar{s}$ & 4.5 & 3.3(27\%) & 2.5(44\%) & 2.3(49\%)     \\
$E$(g/veh/km) & 246.9 & 231.8(6\%) & 226.7(8\%) & 221.3(10\%)   \\
$\bar{w}_s$(m) & 0 & 0 & 0 & 0                  \\
$\bar{w}_a$(m) & 15 & 7.8(48\%) & 7.1(53\%) & 6.7(55\%) \\
\hline
\end{tabular}
\end{center}
\end{table}

Table \ref{tb:eval_modD} presents the evaluation results of a moderate-demand scenario without the freeway incident. The demand change significantly increases the queue measurements ($\bar{w}_s$ and $\bar{w}_a$) compared with table \ref{tb:eval_lowD}. The percentage improvements of each arterial traffic control strategies are close to those in table \ref{tb:eval_lowD} in terms of the travel time, the number of stops and the emission rates. The QL-based control has a strong effect in off-ramp queue ($\bar{w}_s$) dissipation because the off-ramp queue is part of the TSC agent's reward function. The QL-based control also outperforms the MAXBAND with regard to the intersection queue ($\bar{w}_a$), which is taken into account in the DSO agent's reward function.

\begin{table}[h]
\begin{center}
\setlength{\tabcolsep}{6pt} % Default value: 6pt
\caption{Moderate-Demand without Incident} \label{tb:eval_modD}
\begin{tabular}{|c|c|c|c|c|}
\hline
Freeway control & \multicolumn{4}{|c|}{NFC}    \\     \hline
Arterial control & FAC & MAXBAND & QAC & QACU \\    \hline
$T_t^f$(s) & 637 & 639 & 638 & 638              \\
$T_t^a$(s) & 981 & 861(12\%) & 841(14\%) & 834(15\%)    \\
$\bar{s}$ & 4.8 & 3.5(27\%) & 2.9(40\%) & 2.7(44\%)     \\
$E$(g/veh/km) & 254 & 237.2(7\%) & 228.4(10\%) & 226.5(11\%)   \\
$\bar{w}_s$(m) & 52.6 & 46.7(11\%) & 8.8(83\%) & 8.6(84\%)                  \\
$\bar{w}_a$(m) & 47.1 & 34.2(23\%) & 28.4(40\%) & 27.1(42\%) \\
\hline
\end{tabular}
\end{center}
\end{table}

Table \ref{tb:eval_hiD} presents the evaluation results of a high-demand scenario without the freeway incident. The demand increase creates longer queues at off-ramps and intersections that potentially introduces congestion at these areas. Comparing with table \ref{tb:eval_lowD} and \ref{tb:eval_modD}, the percentage improvements brought by the QL-based control diminish in terms of the arterial travel time and the number of stops, closer to those of the MAXBAND. The reason is that the QL-based algorithms prioritize the queue dissipation at off-ramps when it reaches the capacity. They still outperform the MAXBAND in both queue measurements as the MAXBAND does not consider real-time queues at all. Another interesting observation is that the freeway travel time is improved by implementing the QL-based arterial control, because the off-ramp queue does not accumulate and no bottleneck exists on freeway.

\begin{table}[h]
\begin{center}
\setlength{\tabcolsep}{6pt} % Default value: 6pt
\caption{High-Demand without Incident} \label{tb:eval_hiD}
\begin{tabular}{|c|c|c|c|c|}
\hline
Freeway control & \multicolumn{4}{|c|}{NFC}    \\     \hline
Arterial control & FAC & MAXBAND & QAC & QACU \\    \hline
$T_t^f$(s) & 755 & 752 & 704(7\%) & 705(7\%)              \\
$T_t^a$(s) & 1037 & 948(7\%) & 947(7\%) & 940(8\%)    \\
$\bar{s}$ & 6.4 & 5.5(14\%) & 5.6(12\%) & 5.2(19\%)     \\
$E$(g/veh/km) & 272.6 & 250.7(8\%) & 243.8(11\%) & 239.4(12\%)   \\
$\bar{w}_s$(m) & 457.3 & 356.6(22\%) & 82.2(82\%) & 80.1(82\%)                  \\
$\bar{w}_a$(m) & 176.1 & 150.9(14\%) & 83.8(40\%) & 81.5(42\%) \\
\hline
\end{tabular}
\end{center}
\end{table}

Table \ref{tb:eval_lowD_inc} presents the evaluation results of a low-demand scenario with the freeway incident. To alleviate the congestion brought by the incident, we implement an integrated freeway traffic control method proposed in \cite{yuan2023integrated}. Meanwhile we incorporate different arterial traffic control strategies and examine their performance in the occurrence of the freeway control. The percentage improvements of each type of arterial control over FAC in table \ref{tb:eval_lowD_inc} are close to those in table \ref{tb:eval_lowD}, which implies that the freeway control has no significant impact on the arterial performance under low demands. It is also verified by comparing the NFC$+$QACU case with IFC$+$QACU case as they deliver similar performance. 

\begin{table}[h]
\begin{center}
\setlength{\tabcolsep}{6pt} % Default value: 6pt
\caption{Low-Demand with Incident} \label{tb:eval_lowD_inc}
\begin{tabular}{|c|c|c|c|c|c|}
\hline
Freeway control & \multicolumn{4}{|c|}{IFC} & NFC   \\     \hline
Arterial control & FAC & MAXBAND & QAC & QACU & QACU \\    \hline
$T_t^f$(s) & 653 & 653 & 651 & 651 & 730             \\
$T_t^a$(s) & 979 & 852(13\%) & 835(15\%) & 817(17\%) & 807(18\%)    \\
$\bar{s}$ & 4.4 & 3.4(23\%) & 3.2(27\%) & 2.8(36\%) & 2.4(45\%)     \\
$E$(g/veh/km) & 246.1 & 232.1(6\%) & 229.6(7\%) & 225.1(9\%) & 221.9(10\%)   \\
$\bar{w}_s$(m) & 0 & 0 & 0 & 0 & 0                  \\
$\bar{w}_a$(m) & 13.9 & 7.9(43\%) & 7.4(47\%) & 6.6(53\%) & 6.3(55\%) \\
\hline
\end{tabular}
\end{center}
\end{table}

Table \ref{tb:eval_modD_inc} presents the evaluation results of a moderate-demand scenario with the freeway incident. The percentage improvements of each type of arterial control over FAC in this table are close to those in table \ref{tb:eval_modD}. The off-ramp and intersection queue dissipation by QL-based control is also observed under the existence of the freeway incident and the freeway control.

\begin{table}[h]
\begin{center}
\setlength{\tabcolsep}{6pt} % Default value: 6pt
\caption{Moderate-Demand with Incident} \label{tb:eval_modD_inc}
\begin{tabular}{|c|c|c|c|c|c|}
\hline
Freeway control & \multicolumn{4}{|c|}{IFC} & NFC   \\     \hline
Arterial control & FAC & MAXBAND & QAC & QACU & QACU \\    \hline
$T_t^f$(s) & 669 & 670 & 660 & 659 & 745             \\
$T_t^a$(s) & 991 & 887(10\%) & 874(12\%) & 863(13\%) & 872(12\%)    \\
$\bar{s}$ & 5.3 & 3.8(28\%) & 3.5(34\%) & 3.3(38\%) & 3.5(34\%)     \\
$E$(g/veh/km) & 253.9 & 237(7\%) & 232.5(8\%) & 228.6(10\%) & 231.2(9\%)   \\
$\bar{w}_s$(m) & 46.7 & 40.8(13\%) & 5.6(88\%) & 5.3(89\%) & 5.8(88\%)                  \\
$\bar{w}_a$(m) & 49.2 & 39.5(20\%) & 35.3(28\%) & 27.6(44\%) & 34(31\%) \\
\hline
\end{tabular}
\end{center}
\end{table}

Table \ref{tb:eval_hiD_inc} presents the evaluation results of a high-demand scenario with the freeway incident. By implementing the QL-based control, the trade-off between the queue lengths and other performance measurements as mentioned in table \ref{tb:eval_hiD} is observed as well in table \ref{tb:eval_hiD_inc}. The trade-off also leads to a faster travel time on freeway as the off-ramp queue does not accumulate. Under the high demand, the performance of NFC$+$QACU is slightly worse than IFC$+$QACU, which is different from the moderate demand or low demand case. The possible reason is that the incident produces unbalanced off-ramp traffic flows that lead to the incompatibility between the unified signal plan and some intersection demands. The integrated freeway control is effective in balancing off-ramp traffic flows, and thus, improves the performance of QACU.

\begin{table}[h]
\begin{center}
\setlength{\tabcolsep}{6pt} % Default value: 6pt
\caption{High-Demand with Incident} \label{tb:eval_hiD_inc}
\begin{tabular}{|c|c|c|c|c|c|}
\hline
Freeway control & \multicolumn{4}{|c|}{IFC} & NFC   \\     \hline
Arterial control & FAC & MAXBAND & QAC & QACU & QACU \\    \hline
$T_t^f$(s) & 766 & 763 & 726(5\%) & 728(5\%) & 847             \\
$T_t^a$(s) & 1057 & 955(10\%) & 943(11\%) & 932(12\%) & 945(11\%)    \\
$\bar{s}$ & 7.2 & 5.3(26\%) & 4.7(35\%) & 4.3(40\%) & 5.1(29\%)     \\
$E$(g/veh/km) & 273.8 & 247.1(10\%) & 244(11\%) & 239.7(12\%) & 240.6(12\%)   \\
$\bar{w}_s$(m) & 159.4 & 121.9(24\%) & 49.4(69\%) & 45.1(72\%) & 86.7(46\%)                  \\
$\bar{w}_a$(m) & 132.8 & 105.8(20\%) & 66.7(50\%) & 57.8(56\%) & 73.8(44\%) \\
\hline
\end{tabular}
\end{center}
\end{table}

The evaluation results in the above six tables reveal a consistent performance ranking of the listed arterial traffic control strategies, that is QACU, QAC, MAXBAND, FAC from best to worst. Under low traffic demands, the QL-based arterial control produces more benefit than the classic MAXBAND in terms of the travel time, the number of stops and the emission rates at the arterial region. Under high traffic demands, the above-mentioned performance difference is less obvious because the QL agents are designed to prevent queue spillbacks at off-ramps and intersections, while the MAXBAND does not consider that. As a result, the QL-based control reduces the average queue lengths at off-ramps and intersection significantly. Another interesting observation is that the freeway traffic control and the arterial traffic control improves the performance of each other under high traffic demands. The reason is that the arterial control dissipates the off-ramp queue and prevents freeway bottlenecks, and the freeway control balances the off-ramp flow rates and the demands of arterial intersections, which suits the unified signal plan.

\section{Conclusion} \label{section:Conclu}
In this paper, we proposed an arterial traffic control strategy that combines traffic signal control (TSC) and dynamic speed offset (DSO) coordination using a Q-learning (QL) framework. The TSC agent determines the signal cycles and the splits based on intersection demands and off-ramp queue. Then a network controller unifies the cycles and splits of different intersections with a majority rule to facilitate the arterial traffic progression with the assistance of the DSO agent. The DSO agent determines the relative offset and the recommended speeds between two consecutive intersections based on their physical distance, queue lengths and signal cycles. We demonstrated the effectiveness of the proposed approach using microscopic simulations in a mixed freeway and arterial road network with real-world traffic demands. The proposed QL-based control delivers a significantly higher performance than MAXBAND and fixed-time control in terms of travel time and number of stops under low and moderate demands. In high-demand scenarios, the QL-based control trades the travel-time benefit for queue dissipation at off-ramps and intersections, resulting in freeway travel time reduction. The freeway traffic control also improves the performance of the proposed arterial control by providing balanced off-ramp flows in accordance with the unified signal timing.  

Potential extensions of the proposed approach involve a more elaborate discretization of the state and action space, a simpler signal phasing scheme for low-demand scenarios and a more sophisticated QL framework to coordinate freeway and arterial traffic control.

%
% ---- Bibliography ----
%
% BibTeX users should specify bibliography style 'splncs04'.
% References will then be sorted and formatted in the correct style.
%
% \bibliographystyle{unsrt}
% \bibliography{biblio}
%

\end{document}